\documentstyle[preprint,prl,aps]{revtex}

\input BoxedEPS
\SetRokickiEPSFSpecial  
\HideDisplacementBoxes

\begin{document}

\draft
\tighten
\preprint{MIT-CTP 2780\quad nucl-th/9809063}
\title{Superscaling in inclusive electron-nucleus scattering}
\author{T.W. Donnelly }
\address{
Center for Theoretical Physics,  Laboratory for Nuclear Science\\
and Department of Physics\\
Massachusetts Institute of Technology\\
Cambridge, Massachusetts 02139, USA
}
\author{Ingo Sick}
\address{
Departement f\"ur Physik und Astronomie,
Universit\"at Basel\\ 
CH-4056 Basel, Switzerland 
}
\date{September 18, 1998}
\maketitle
\begin{abstract}We investigate the degree to which the scaling functions
$F(\psi ')$  derived from cross sections for inclusive electron-nucleus 
quasi-elastic scattering 
define the {\em same} function for {\em different} nuclei. In the region 
where the scaling variable $\psi ' < 0$, we find that this superscaling
is experimentally realized to a high degree.
\end{abstract}

\section{Introduction}
The use of scaling and the application of dimensional analysis to
cross sections for inclusive scattering of a weakly interacting probe
from the constituents of a composite system have been
important tools for the development of new insights in physics.
Examples for such processes are  the scattering of keV electrons from 
electrons bound in atoms \cite{Bonham77},  the scattering of eV
neutrons from atoms in solids or liquids \cite{Hohenberg66}, deep inelastic
scattering of GeV energy leptons from the quarks in the nucleon \cite{Bjorken69}
 and, of particular interest for the present letter, quasi-elastic 
scattering of electrons in the energy range of 100's of MeV to several GeV
from nucleons in nuclei \cite{Sick80}.  Despite the 
extraordinary range of energy and momentum transfer for which scaling has
been studied, the conceptual basis for describing this phenomenon has many 
features in common.

 The inclusive cross sections for these processes in general depend 
 explicitly on two independent variables -- the energy $\omega$ and momentum  $\vec{q}$ transferred by the probe to the 
constituent. {\em Scaling} means that, in the asymptotic regime of large $q$ 
and $\omega$, the 
cross sections depend on a {\em single} variable $z = z(q,\omega)$, 
itself a function of $\omega$ and $q$. This property
results essentially from the kinematics of the scattering process of
the probe by the moving constituent, which as a consequence of the recoil 
momentum received, is ejected essentially quasi-freely from the composite 
system.  

The interest in scaling phenomena originates from two distinct sources: 
\begin{itemize}
\item The observation of the occurrence (or nonoccurrence) of scaling yields 
information on the domination (or not) of the quasi-free scattering process 
or the contribution 
of other reaction mechanisms (which in general do not scale). This provides 
{\em experimental} insight into the reaction mechanism, knowledge which is a 
prerequisite for a quantitative understanding  of the cross section. 
\item The function to which the data scale is closely related to the momentum 
distribution 
of the  constituents in the composite target. This provides interesting 
experimental knowledge on the dynamics of the bound system.
\end{itemize}

For quasi-elastic electron-nucleus scattering, data are available over a 
large range of $(q,\omega)$ and for several nuclei. These data have been found 
to scale in a major 
part of the kinematical region studied. When analyzed in terms of the scaling 
variable $y$ \cite{Sick80}, which is close to the component  $k_{//}$ of a
bound nucleon's
momentum parallel to $\vec{q}$, the data exhibit scaling for $y<0$
(roughly $ \omega <|Q^2|/2m_N$, with $m_N$ the nucleon mass, where $Q^2=\omega^2
-q^2 <0$), that is, the region in which quasi-elastic 
scattering dominates. Much of the past work has concentrated 
on the study of the scaling properties of the response in the low-$\omega$ 
tail of the quasi-elastic peak where $y$ is large and negative. Detailed 
quantitative studies of the conditions under which scaling occurs and the 
impact of adverse effects such as final-state interactions 
\cite{Benhar91}, the spread of the spectral function $S(k,E)$ in removal 
energy $E$, and the contributions 
of other reaction mechanisms have been made. For a review see \cite{Day90}. 

Past applications of scaling focused on individual nuclei and nuclear matter 
using the 
data for different kinematics in the range $q$ = 0.5 -- 2 GeV/c and $\omega$ = 
0.1 -- 3 GeV. In the present letter, we explore a novel aspect: rather than
concentrating on the response of individual nuclei, we compare the 
scaling function of {\em different nuclei} with mass number $A\geq$4, and 
study the degree to which these scaling functions are the {\em same}. 
Our goal is to explore whether 
the concept of superscaling, introduced in  \cite{Alberico88} within the 
relativistic Fermi gas (RFG) model,  works for finite nuclei. For these 
studies we concentrate on the main part of the 
quasi-elastic peak where superscaling could, if anyplace, be hoped to work.

\section{Formalism} 
Discussions of scaling at intermediate energies assume
that inclusive electron scattering in the quasi-elastic
regime has as a dominant process the impulsive one-body knockout of
nucleons together with contributions from two-body meson exchange current 
(MEC) and meson production processes that may play a role when the normally 
dominant process is 
suppressed. In addition to the electron scattering angle $\theta_e$, one
has two variables to characterize the cross section, typically chosen to
be $(q,\omega)$. Of course, any function
$z=z(q,\omega)$ of $q$ and $\omega$ may be used together with $q$ to
characterize the cross section. For over two decades it has been 
traditional to use the so-called $y$-scaling variable for $z(q,\omega)$
(for a review of the history of the subject, the basic formalism and
reference to many of the theoretical studies undertaken see \cite{Day90}).
Upon dividing the inclusive electron scattering cross section by the 
single-nucleon electromagnetic cross section together with the 
Jacobian required in changing variables to $y$ one obtains a derived
function $F(q,y)$. Scaling means that at high enough values of momentum
transfer this function becomes a function only of $y$, that is, independent
of $q$:
\begin{equation}
F(q,y) \stackrel{q \to \infty}\longrightarrow F(y)\equiv F(\infty,y) .
\label{scaling}
\end{equation}
Indeed, it has been found that in the $y<0$ region (the region of
the low-$\omega$ side of the quasi-elastic peak) for momentum transfers
of roughly 0.5 GeV/c or larger $y$-scaling is very well obeyed. For more
detailed discussion of this approach and for a summary of the experimental
situation, see \cite{Day90}.

As an alternative to this standard plane-wave impulse approximation 
approach scaling can be examined from a different point of view using
as a starting point the RFG model \cite{Alberico88,Barbaro98}.
This model has the appeal of simplicity while maintaining important aspects
in the problem such as Lorentz covariance and gauge invariance. Naturally
it ignores potentially important effects such as those stemming from strong
final-state interactions or two-body MEC and employs an overly simplified
initial-state spectral function; nevertheless, such ingredients can be
added to the basic model and appear not to invalidate it as a basic
starting point for analyses of scaling. 

Below we summarize the essential
RFG developments and refer the reader to \cite{Barbaro98,Donnelly98a} for
more detail, including the relationships between the RFG formalism and
the usual $y$-scaling analysis. As seen in the work cited,
a dimensionless scaling variable $\psi$ naturally emerges:
\begin{equation}
  \psi = \frac{1}{\sqrt{\xi_F}} \frac{\lambda - \tau}
   {\sqrt{(1+\lambda)\tau + \kappa \sqrt{\tau (1+\tau)}}} 
\label{psi}
\end {equation}
where $\xi_F = \sqrt{1 + \eta_F^2} - 1$ and $\eta_F =
k_F/m_N$ are the dimensionless Fermi kinetic energy and momentum, respectively.
Here we employ dimensionless variables $\kappa\equiv q/2m_N$, 
$\lambda\equiv \omega/2m_N$ and $\tau\equiv\kappa^2-\lambda^2>0$.
Additionally, to allow for the fact that nucleons are
knocked out of all shells in the nucleus we follow the spirit of 
\cite{Cenni97} and 
introduce an ``average removal energy'' by shifting $\omega$ to $\omega'\equiv
\omega - E_{rem}$ and hence defining a derived variable $\psi'$ by
substituting $\lambda'\equiv \omega'/2m_N$ for $\lambda$ and 
$\tau'\equiv \kappa^2-{\lambda'}^2$ for $\tau$ in 
Eq.~(\ref{psi}). In fact, in the present work we simply adjust $E_{rem}$
to fit the systematic trend seen in the data, rather than employing the
specific shift discussed in \cite{Cenni97}; however, since typically 
$\omega$ is large for the kinematics considered below, while the chosen 
values of $E_{rem}$ are small, ranging from about 15 to 25
MeV in our analysis, the actual values chosen are not critical.

Weighting the squares of the single-nucleon form factors with the appropriate
proton and neutron numbers, $Z$ and $N$, 
\begin{eqnarray}
  {\widetilde G}_E^2 (\tau) &\equiv& Z G_{Ep}^2 + N G_{En}^2 \nonumber \\
  {\widetilde G}_M^2 (\tau) &\equiv& Z G_{Mp}^2 + N G_{Mn}^2 
\label{formfac}
\end{eqnarray}
within the RFG model it is then natural to define a function
\begin{equation}
F (\kappa, \psi') \cong \frac{k_F\, d^2 \sigma/d \Omega_e\, d \omega}
   {\sigma_M [v_L (\kappa/2\tau) {\widetilde G}_E^2 
    + v_T (\tau/\kappa) {\widetilde G}_M^2 ]} 
\label{Ftotal}
\end{equation}
where terms of order $\eta_F^2$ have been ignored here
since $\eta_F$ is typically small, growing from 0.06 for deuterium 
to about 0.3 for the heaviest nuclei (full expressions are given in
\cite{Barbaro98,Donnelly98a}). Here $\sigma_M$ is the Mott cross section
and $v_{L,T}$ the familiar Rosenbluth kinematical factors.
In terms of the RFG variables scaling
means that at high $q$ (large $\kappa$) one finds that $F (\kappa,\psi')$
becomes only a function of $\psi'$ -- indeed, if $E_{rem}$ is set to
zero, then the RFG scales (in $\psi$) exactly, as it must, since this
provided the original definition of $\psi$ \cite{Alberico88}.

While the RFG approach discussed here and the usual $y$-scaling analysis
have differences in detail, they are rather closely related under
``typical'' circumstances, namely, for momentum transfers that are
relatively large (say $q>$ 0.5 GeV/c), for energy transfers that are
not too close to threshold (where both approaches are bound to fail) and
for nuclei that are not too light (since the RFG approach cannot account 
for daughter nucleus recoil effects). Roughly speaking the $\psi'$
variable can be regarded as an alternative to $y/k_F$ and $F(\kappa,\psi')$
can be equated with $k_F \times F(q,y)$. One finds that scaling in 
$\psi'$  and $y$ is very similar (this is discussed in more detail in a
longer accompanying paper \cite{Donnelly98a}).

In \cite{Alberico88} the idea of ``superscaling'' was introduced, but until
now not tested with existing high-quality data on
quasi-elastic scattering. The observation to be drawn from the RFG developments
summarized above is that, within the context of that model,
plotting $F (\kappa,\psi)$ versus $\psi$ leads to a universal curve, namely,
not only universal in that no $\kappa$-dependence remains, but also that
the result {\em is independent of the choice of nucleus} modulo minor 
corrections 
of order $\eta_F^2$. This suggests attempting the same type of representation 
of the experimental data to investigate the idea of superscaling, which 
we proceed to do in the following section.

\section{Results} 
We next use the inclusive electron-nucleus scattering data 
in order to test the idea of superscaling. We also try to 
disentangle better the various reasons that underlie the well-known fact that at
large electron energy loss nonscaling is observed. For these studies we 
basically concentrate on nuclei with $A\geq 4$, as the lightest nuclei are 
known to have momentum distributions that are very far from the 
``universal'' one which is at the basis of the superscaling idea.

Data on inclusive electron-nucleus scattering for a series of nuclei 
$A \geq 4$ are available  in the region of low momentum transfers 
$q \sim$ 0.5 GeV/c  
\cite{Whitney74} --\nocite{Barreau81,Barreau83,Baran88,Deady86,Meziani85,Yates93}
\nocite{Williamson97,Altemus80,Meziani85,Zghiche94,Chen91,Connell87}
\nocite{Titov72}\cite{Blatchley86}, while 
data extending to much  higher $q$  are available from other experiments  
\cite{Day93} --\nocite{Heimlich74,Day89a}\cite{Hotta84}. Not 
all of these data can be used, however, as some of them have not been 
corrected for radiative and Coulomb distortion effects, are known to have 
problems such as ``snout scattering'' or the inclusion of false signals
from $\pi^-$'s in the electron spectrometer, 
or are only available in the form of figures. Part of the data is at very
low momentum transfer, which we do not consider as there scaling is known to
break down due to large final-state interactions and Pauli blocking effects.

In a first step, we have taken the data that meet our criteria 
for the nuclei $A =  12\ldots 208$
and have analyzed them in terms of scaling in the variable $\psi'$. As $\psi'$
is defined in terms of the Fermi momentum, appropriate values of $k_F$ had to
be selected. We use 220, 230, 235 and 240 MeV/c for C, Al, Fe, Au, with 
intermediate
values for the intermediate nuclei. 

\begin{figure}[htb]
\vspace*{-1pc}
$$ \BoxedEPSF{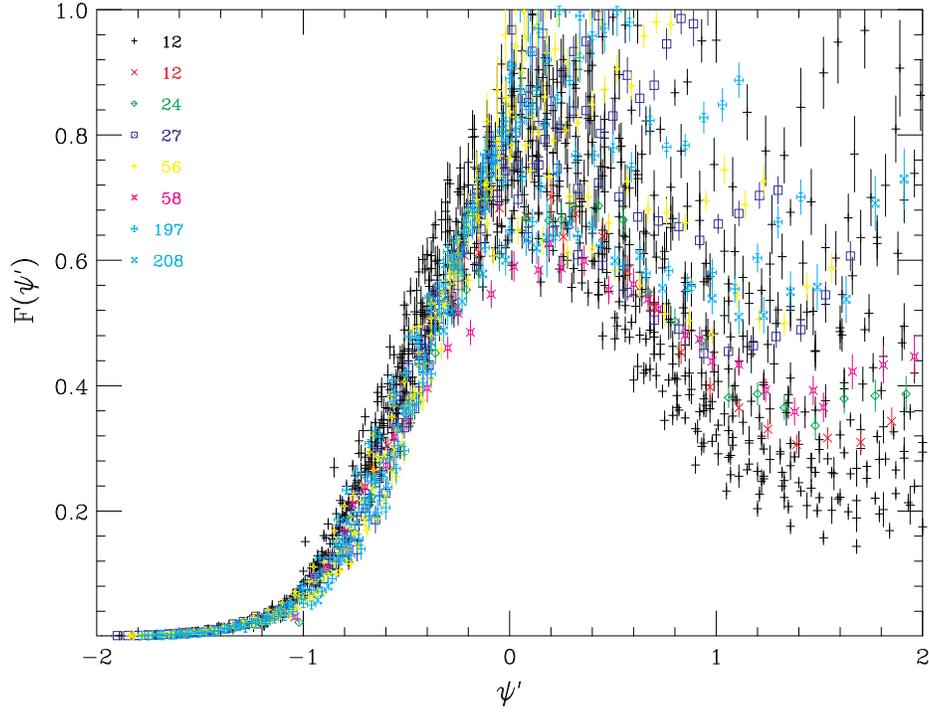 scaled 750}$$
 \caption{Scaling function $F(\psi')$ as function of $\psi'$ for all nuclei
\protect{$A \ge $ 12} and all kinematics. The values of A corresponding to 
different symbols is shown in the insert. 
} \label{di421} 
 \end{figure}

Figure~\ref{di421} shows the scaling function $F(\psi')$ for all kinematics
(energies, angles, momentum transfers) suitable for the present study
and all $A$ available.  We clearly observe a
scaling behavior for values of $\psi ' < 0$: while the cross 
sections at a given $\psi'$ vary over more than three orders of magnitude, 
the values of
$F(\psi ')$ are essentially universal.  For $\psi ' >0$, on the other hand,  
the scaling property is badly violated, and this is to be expected, as there 
processes other 
than quasi-elastic scattering -- meson exchange currents, $\Delta$-excitation,
deep inelastic scattering -- contribute to the cross section. The scaling
as discussed in this paper applies only to processes having the kinematics of 
electron-nucleon quasi-free scattering.

\begin{figure}[htb]
\vspace*{-1pc}
$$ \BoxedEPSF{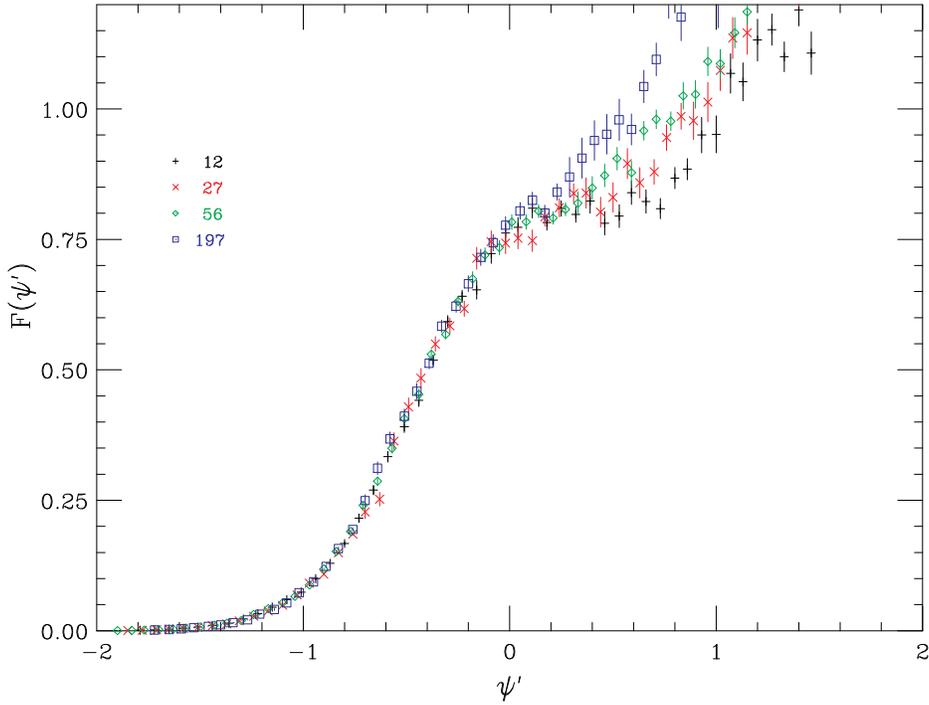 scaled 750}$$
\caption{Scaling function for C, Al, Fe, Au and fixed kinematics 
\protect\cite{Day93}. 
The correspondence of symbol and mass number of the nucleus is shown in the 
insert.
}
\label{di321}  
\end{figure}

In order to separate some of the effects leading to less-than-perfect 
scaling at negative 
$\psi'$, in Fig.~\ref{di321} we show the function $F(\psi') $ for the 
series of 
nuclei $A = 12\ldots 197$, but for fixed kinematics (electron energy 3.6 GeV, 
scattering angle 16$^\circ$, and hence nearly constant $q$). The quality of the 
scaling in the region 
$\psi ' < 0 $ is quite amazing. This shows that, insofar as the removal of 
the $A$-dependence goes, the superscaling works extremely well. The deviations
from scaling observed in Fig.~\ref{di421} are {\em not\/} from an $A$-dependence.

A part of the $A$-dependent increase of $F(\psi')$ at positive $\psi'$ results 
from 
the increase of $k_F$ with $A$. This amounts to an increase of the width of the 
quasi-elastic and $\Delta$ peaks, and a correspondingly increased overlap with 
non--quasi-free scattering processes ($\Delta$-excitation, $\pi$-production, $\ldots$). 
At the same time, the increasing average density of 
the heavier nuclei may also lead to an increase in contributions of two-body
MEC processes which are presumably strongly density-dependent
(i.e., do not scale with $k_F$ in the same way the one-body knockout
processes do; see \cite{Vanorden81} for indications of this type of 
behavior).   

\begin{figure}[htb]
\vspace*{-1pc}
$$\BoxedEPSF{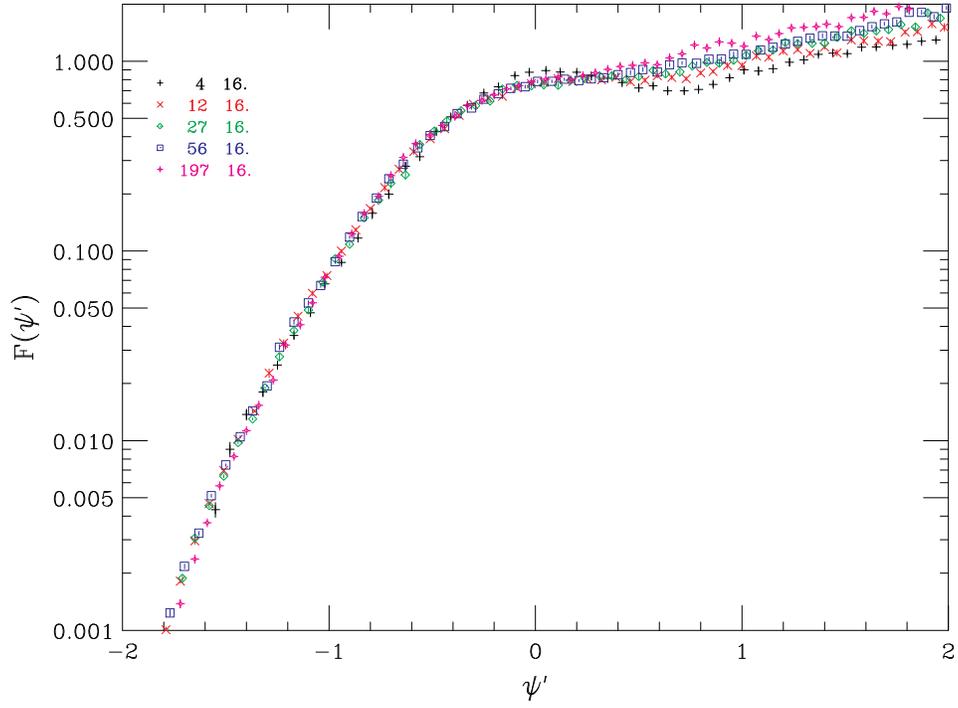 scaled 750}$$
 \caption{Scaling function for nuclei $A = 4$ -- 197  and fixed 
kinematics  on logarithmic scale. 
}\label{di422} 
\end{figure}

Figure \ref{di422} shows the data for $A=$ 4, 12, 27, 56, 197 
on a logarithmic scale for the kinematics of Fig.~\ref{di321}. 
This figure demonstrates that 
the scaling property extends to large negative values of $\psi'$, values which
correspond to large momenta of the initial nucleon. A priori, this feature is 
not  predicted within the RFG model used to motivate the 
choice of $\psi'$. It can be understood, however, from the theoretical 
results for the momentum distribution of nuclear matter as a function of 
the nuclear matter density. For different nuclear matter densities,
the tail of  the momentum distribution $n(k)$ at $k > k_F$ (corresponding to 
$\psi ' < -1$) is a near-universal function of $k/k_F$ 
\cite{Baldo90}. For finite nuclei and large momenta we can employ the Local 
Density  Approximation (LDA), as at large $k$ we are dealing with short-range 
properties of the nuclear wave function \cite{Benhar94b}. Within LDA, the 
nuclear momentum distribution (spectral function) then is a weighted  average
over the corresponding nuclear matter distributions. This means that 
the large-momentum tail of the nuclear momentum distribution also scales with
$k_F$, a dependence that is removed when using $\psi'$.

In order to emphasize the quality of this superscaling in the tail, 
in  Fig.~\ref{di422} we have also included the data on $^4$He, taken under 
the same kinematical conditions \cite{Day93}.
While at $\psi '$=0  $F(\psi ')$ for $^4$He 
is about 10\% higher than for heavier 
nuclei as a consequence of the  sharper peak of the momentum distribution $n(k)$
at $k \sim 0$, the scaling function 
for $^4$He agrees perfectly with the one for heavier nuclei for 
$\psi ' < - 0.2$.

\begin{figure}[htb]
\vspace*{-1pc}
$$\BoxedEPSF{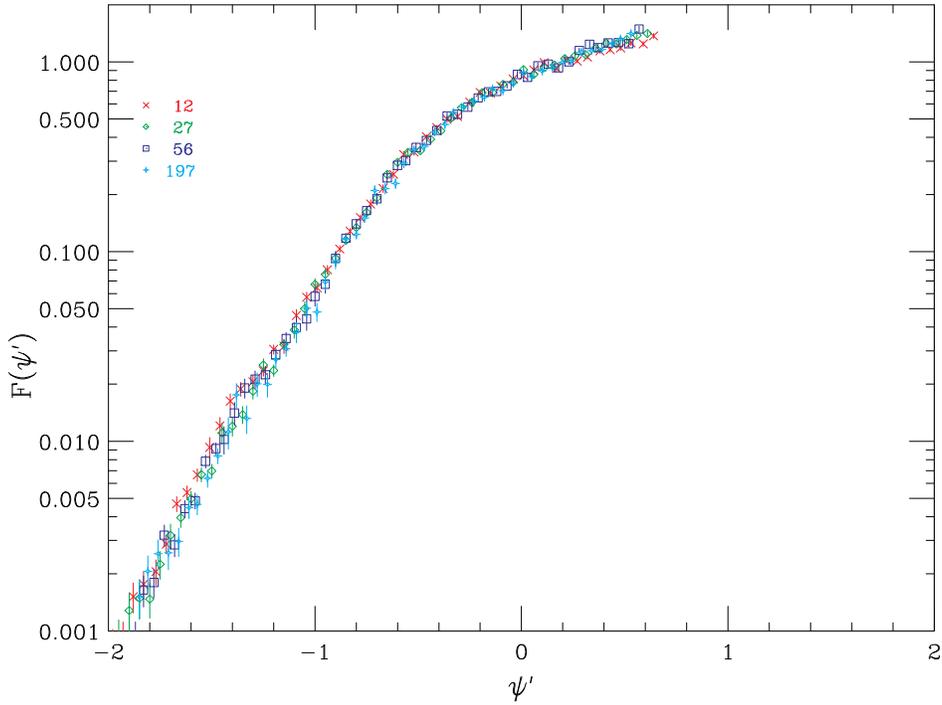 scaled 750}$$
 \caption{Scaling function for nuclei $A$ = 4 -- 197  at higher momentum
transfers (3.6 GeV, 25$^\circ$).
}\label{di47} 
\end{figure}

A similar quality of superscaling is found when analyzing the data  
at the other kinematics at both higher and lower $q$ where cross sections for 
a large range of $A$ are available. As an example, Fig.~\ref{di47} shows
the scaling function for the data at 3.6 GeV and 25$^\circ$ \cite{Day93} 
  
The quality of scaling in terms of the variable $\psi'$ is quite similar to
the one found when using the standard variable $y$. We  have systematically 
used $\psi'$ defined above as this variable is more directly connected to the 
RFG model that motivated the study of superscaling.
  
\section{Conclusions} We have analyzed data on 
electron-nucleus quasi-elastic  scattering for nuclei with mass numbers 
$A$ = 4  -- 208 which cover a large range in $q,\omega$. We find that, upon 
use of the scaling variable $\psi'$ 
which allows one to remove the ``trivial'' dependence on the Fermi 
momentum, the data on the low-$\omega$ side of the quasi-elastic
peak ($\psi ' < 0$)  show superscaling, {\em i.e.} the scaling functions
 $F(\psi ')$ for the different nuclear mass numbers $A$ coincide. 
The $A$-independence of the  
superscaling function actually is much better realized than the 
$q$-independence of the normal scaling function, which 
is violated due to processes such as meson exchange currents and/or excitation 
of internal  degrees of freedom of the nucleon. The realization of superscaling
shows that different nuclei, in the integral sense tested via inclusive 
scattering \cite{Day90}, have a more or less universal momentum distribution 
once the obvious  dependence on the Fermi momentum $k_F$ is removed. 

\acknowledgments
We would like to thank J. Jourdan and 
C.F. Williamson 
for help and useful discussions. This work was supported in part by funds 
provided by the U.S. Department  of Energy under cooperative research agreement
\#DF-FC02-94ER40818, and by the Swiss National Science Foundation.  

\end{document}